\documentclass[twocolumn,showpacs,preprintnumbers,amsmath,amssymb]{revtex4}

\usepackage{graphicx}
\usepackage{dcolumn}
\usepackage{bm}

\graphicspath{%
    {converted_graphics/}
    {C:/Users/Nail/Desktop/}
    {C:/Users/Nail/Desktop/Mathematica/}
}
\begin{document}

\preprint{IUHET-542}

\title{Isospectral Potentials from Modified Factorization}

\author{Micheal S. Berger}
 \email{berger@indiana.edu}
\author{Nail S. Ussembayev}%
 \email{nussemba@indiana.edu}
\affiliation{%
Department of Physics, Indiana University, Bloomington, IN 47405, USA}%

\begin{abstract} 
Factorization of quantum mechanical potentials has a long history extending back to the earliest days of the subject. In the present paper, the non-uniqueness of the factorization is exploited to derive new isospectral non-singular potentials. Many one-parameter families of potentials can be generated from
known potentials using a factorization that involves superpotentials defined in terms of excited states of a potential. For these cases an operator representation is available. If ladder operators are known for the original potential, then a straightforward procedure exists for defining such operators for its isospectral partners. The generality of the method is illustrated with a number of examples which may have many possible applications in atomic and molecular physics. 

\end{abstract}
\pacs{03.65.Ge, 03.65.Fd, 03.65.Ca}
      
\maketitle

\section{Introduction}       
The factorization method due to Hull and Infeld \cite{Hull}   has been widely exploited in quantum mechanics 
to determine the spectra and wave functions of exactly solvable potentials.
This approach has been formalized in supersymmetric quantum mechanics 
(SUSY QM) \cite{Cooper} which has been used to find many new isospectral 
potentials. The usual procedure is to find a factorization of a quantum 
mechanical Hamiltonian and the methods of SUSY QM then guarantee that a 
supersymmetric partner potential is isospectral to the original Hamiltonian. 
As verified below, this procedure yields a pair of potentials with the same spectra (possibly apart from the ground state) and related wave functions. Throughout this paper we work in $\hbar=2m=1$ units.

Let's consider a one dimensional Hamiltonian \[H_-^{(0)}=-\partial^2_x+V_-^{(0)}(x)\] where $V_-^{(0)}(x)$ is an arbitrary non-singular potential with at least one bound state and zero ground state energy (given the Hamiltonian $H=-\partial^2_x+V(x)$ one simply subtracts the zero point energy to obtain $H_-^{(0)}$). It is a second order linear operator  and it  can be factored into a product of first order linear operators as follows: \[H_-^{(0)}=(-\partial_x+W_0(x))(\partial_x+W_0(x))\equiv A_0^\dag A_0\] once the ground state wave function $\psi_0(x)$ is specified. The function $W_0(x)=-\partial_x\ln\psi_0(x)$ is called superpotential generating the potential \[V_-^{(0)}(x)=W_0^2(x)-W_0'(x).\]  

Fortunately,  the factorization does not commute $A_0^\dag A_0\ne A_0A_0^\dag$ unless the superpotential is constant. In other words, an inverted product $A_0A_0^\dag$ is a certain new Hamiltonian $H_+^{(0)}=A_0A_0^\dag=-\partial^2_x+V_+^{(0)}(x)$ where \[V_+^{(0)}(x)=W_0^2(x)+W_0'(x)\] is also free of singularities. It turns out that the eigenfunctions and eigenvalues of these partner Hamiltonians are related. Indeed, we have the following first-order intertwining relations \begin{equation}H_-^{(0)}A_0^\dag=A_0^\dag H_+^{(0)} \mbox{ and } H_+^{(0)}A_0=A_0H_-^{(0)} \end{equation} from which one observes that since $A_0\psi_0(x)=0$, the spectra of $H_+^{(0)}$ and $H_-^{(0)}$ are connected by $\tilde E_n=E_{n+1}$ $(n=0,1,\dots)$ where $\tilde E_n$ and $E_n$ denote the eigenvalues of the Hamiltonians $H_+^{(0)}$ and $H_-^{(0)}$ respectively with eigenfunctions $\tilde\psi_n$ and $\psi_n$. Thus, the Hamiltonians have identical energy spectrum except for the ground state of $H_-^{(0)}$. The wave functions satisfy $\tilde\psi_n(x)\propto A_0\psi_{n+1}(x)$, $\psi_{n+1}(x)\propto A_0^\dag\tilde\psi_n(x)$ and if $\psi_{n+1}(x)$ is normalizable, then $\tilde\psi_n(x)$ is also normalizable and vice versa, because \begin{eqnarray*}\langle\tilde\psi_n(x), \tilde\psi_n(x)\rangle&=&\langle\psi_{n+1}(x), A_0^\dag A_0\psi_{n+1}(x)\rangle\\ &=&E_{n+1}\langle\psi_{n+1}(x), \psi_{n+1}(x)\rangle.\end{eqnarray*} Note that for singular potentials (for instance, with a $1/x^2$ singularity) some of the wave functions $\tilde\psi_n(x)$ are not acceptable as they may not be normalizable \cite{Berger}. That is, for singular potentials the degeneracy of energy levels is only partially valid or invalid at all. The upshot of all this is that one can generate new isospectral potentials from existing exactly solvable potentials.   

Luckily, the above discussed factorization is not unique. For example, we have \[(-\partial_x+1)(\partial_x+1)=(-\partial_x+\tanh(x))(\partial_x+\tanh(x)),\] i.e. two different superpotentials can give rise to the same potential (in this particular example with no bound states). One can try to construct new isospectral potentials exploiting non-uniqueness of factorization and obtain a one-parameter family of potentials with the parameter arising as an integration constant  \cite{Mielnik, Mitra}. 

 Suppose the Hamiltonian $H_+^{(0)}$ can be factorized by the operators different than $A_0$ and $A_0^\dag$, namely, \[B=\partial_x +f(x)\mbox{ and } B^\dag=-\partial_x +f(x)\] where $f(x)$ is temporarily undetermined function:
 \[H_+^{(0)}=BB^\dag=-\partial^2_x+f^2(x)+f'(x).\] Now demanding that this Hamiltonian involve the potential $V_+^{(0)}(x)$ results in 
a differential equation that must be satisfied \[f'(x)+f^2(x)-V_+^{(0)}(x)=0.\] This is a Riccati equation in its canonical form.  The explicit closed-form solution of this equation is not known typically, but one understands that the superpotential $W_0(x)$ is a particular solution. This is enough to construct the general solution $f(x)$ which depends on an arbitrary integration constant that can be considered as a free parameter in the partner Hamiltonian \[H=B^\dag B=-\partial^2_x+ V_+^{(0)}(x)-2f'(x)=-\partial^2_x+ V(x).\]
According to SUSY QM the potentials $V_+^{(0)}(x)$ and $V(x)$ are isospectral (except for the lowest state of $V(x)$) provided that $f(x)$ is nonsingular. In addition, since $BB^\dag=A_0A_0^\dag$, it follows that the potentials $V_-^{(0)}(x)$ and $V(x)$ have strictly identical spectra.  

In ref.~\cite{Mielnik} Mielnik performed factorization of the harmonic oscillator potential in this manner. Mielnik obtained one-parameter family of potentials with the oscillator spectrum, but as we have just seen the procedure is straightforwardly 
generalized to any potential $V_+^{(0)}(x)$.  
  
In the standard (i.e. based on the first-order intertwining relation (1)) unbroken SUSY QM it is impossible to use an excited state of the original potential and at the same time avoid creating singularities in the partner potential  \cite{Panigrahi}. There is no guarantee that the resulting wave functions are normalizable and energy levels degenerate.  The purpose of the present article is to modify the operators $B$ and $B^\dag$ in such a way as to determine new strictly isospectral potentials  without being forced to solve Riccati equations (by reducing  the Riccati equation whose appearance in the factorization problems is typical to the solvable Bernoulli equation) and, more importantly,  by applying the non-uniqueness of factorization to the superpotentials generated by the excited states of a potential, since these also satisfy the Schr\"odinger equation.

\section{Modified factorization}
In this section we show the consequences of the non-uniqueness of factorization method extended to the excited states of a potential, rather than just the ground state. In the literature the Hamiltonians $H_+^{(0)}$ and $H_-^{(0)}$ are called "bosonic" and "fermionic" respectively. We show that the degeneracy of energy levels of partner potentials depends on whether the bosonic or fermionic Hamiltonians admit non-unique factorization. 
\subsection{Bosonic Hamiltonian}
Let there be given an analytically solvable non-singular potential $V_-^{(0)}(x)$ whose energy eigenvalues $E_n$ and wave functions $\psi_n(x)$ are known. Without loss of generality, let $E_0$ be zero, so that $V_-^{(0)}(x)=\psi_0''(x)/\psi_0(x)=W_0^2(x)-W_0'(x)$ and also define \[V_-^{(n)}(x)=\psi_n ''(x)/\psi_n(x)=W_n^2(x)-W_n'(x)\] where $W_n(x)=-\partial_x\ln\psi_n(x)$ is taken to be the superpotential corresponding to $\psi_n(x)$. From the Schr\"odinger equation it follows that $V_-^{(n)}(x)=V_-^{(0)}(x)-E_n$, so that the potentials $V_-^{(n)}(x)$ are non-singular, even though the superpotentials $W_n(x)$ are always singular for $n>0$. Adjusting the energy scale seems appropriate: one simply subtracts from the potential the energy of the excited state so that the resulting potential can be factored. 

Next we introduce the operators
\[B_n=\partial_x +f(x)+W_n(x)\mbox{ and } B_n^\dag=-\partial_x+f(x)+W_n(x)\] where $f(x)$ will be determined below. Notice when $n=0$ these definitions reduce to the familiar case of standard unbroken SUSY QM if $f(x)=0$ and to the Mielnik's factorization \cite{Mielnik} if $f(x)\ne 0$. 

The factorization of the Hamiltonian $\tilde H_-^{(n)}=B_n^\dag B_n$ leads to 

\[\tilde H_-^{(n)}=-\partial^2_x+V_-^{(n)}(x)+f^2(x)+2W_n(x) f(x)-f'(x).\] If we require that $f^2(x)+2W_n(x) f(x)-f'(x)=0$ the Hamiltonian becomes trivial because the potential $V_-^{(n)}(x)$ is related to $V_-^{(0)}(x)$ by a constant shift. On the other hand, the partner Hamiltonian $\tilde H_+^{(n)}=B_nB_n^\dag$ is less trivial \[\tilde H_+^{(n)}=-\partial^2_x+V_+^{(n)}+2f'(x)\] where $V_+^{(n)}(x)=W_n^2(x)+W_n'(x)$.
The function $f(x)$ is not arbitrary -- it is a solution of the Bernoulli equation (a specific example of the Riccati equation):
\[f'(x)=f^2(x)+2W_n(x) f(x)\] and reads \[f_n(x)=\frac{\psi^{-2}_n(x)}{C-\int_{x_0}^x\psi^{-2}_n(s)ds}\] where $C$, $x_0$ are constants. It follows that $\psi_n(x)$ must be inverse square integrable; however, in general the wave functions do not possess this property. 

There is yet another problem, namely, singularity of the potentials $V_+^{(n)}(x)$ for $n\ne0$ corresponding to the zeros of the wave functions.  Consequently, the breakdown of the degeneracy of energy levels of the Hamiltonians $\tilde H_-^{(n)}$ and $\tilde H_+^{(n)}$ occurs (in addition to $H_-^{(n)}$ and $H_+^{(n)}$). 

\subsection{Fermionic Hamiltonian}
The difficulties of establishing the degeneracy theorem for bosonic Hamiltonians suggest to reverse the order of the operators $B_n$ and $B_n^\dag$ and start with the fermionic Hamiltonian $\tilde H_+^{(n)}=B_nB^\dag_n$:

\[\tilde H_+^{(n)}=-\partial^2_x+V_+^{(n)}(x)+f^2(x)+2W_n(x) f(x)+f'(x)\] where $V_\pm^{(n)}(x)$ are defined as usual. We again obtain the Bernoulli equation \[f'(x)+f^2(x)+2f(x)W_n(x)=0\] whose general solution is \begin{equation}f_n(x)=\frac{\psi^2_n(x)}{C+\int_{x_0}^x\psi^2_n(s)ds}\end{equation} where $C$, $x_0$ are constants and $\psi_n(x)$ is assumed to be square-integrable. 

If it is possible to restrict the domain of the parameter $C$ and make $f_n(x)$ free of singularities, then  the potential $\tilde V_-^{(n)}(x)$ in \[\tilde H_-^{(n)}=B_n^\dag B_n=-\partial^2_x+\tilde V_-^{(n)}=-\partial^2_x+V_-^{(n)}-2f_n'(x)\] constitute a one-parameter family of potentials isospectral to the potential $V_-^{(n)}(x)$. 

To see this note that the Schr\"odinger equation $H_-^{(n)}\psi_k=(E_k-E_n)\psi_k$ implies \begin{eqnarray*}\tilde H_-^{(n)}[B_{n}^\dag A_n\psi_k]&=&B_{n}^\dag B_{n}B_{n}^\dag A_n\psi_k\\ &=& B_{n}^\dag A_nA_n^\dag A_n\psi_k\\&=&(E_k-E_n)[B_{n}^\dag A_n\psi_k]\end{eqnarray*} where we have used the non-uniqueness of factorization of the Hamiltonian $H_+^{(n)}=A_nA_n^\dag=B_nB_n^\dag$. So if $\psi_k(x)$ is an eigenfunction of the Hamiltonian $H_-^{(n)}$ with energy eigenvalue $E_k-E_n$, then $B_{n}^\dag A_n\psi_k$ is an eigenfunction of $\tilde H_-^{(n)}$ with the same energy.  
Similarly, from the Schr\"odinger equation $\tilde H_-^{(n)}\tilde\psi_k^{(n)}=\tilde E_k^{(n)}\tilde\psi_k^{(n)}$ (where in $\tilde E_k^{(n)}$, $k$ denotes the energy level and $(n)$ refers to the $n^{th}$ eigenfunction of the Hamiltonian $H_-^{(n)}$) it follows that 

\[H_-^{(n)}[A_n^\dag B_{n}\tilde\psi_k^{(n)}]=\tilde E_k^{(n)}[A_n^\dag B_{n}\tilde\psi_k^{(n)}]\]

Hence, the normalized eigenfunctions of the Hamiltonians $H_-^{(n)}$ and $\tilde H_-^{(n)}$ are related by \begin{equation}\tilde\psi_k^{(n)}(x)=(E_k-E_n)^{-1}[ B_{n}^\dag A_n\psi_k(x)]\end{equation} and \[\psi_k(x)=(E_k-E_n)^{-1}[ A_n^\dag B_{n} \tilde\psi_k^{(n)}(x)]\] where $k\ne n$. The operators $A_n$ or $B_{n}$ destroy a node in the eigenfunctions, but they are followed respectively by the operators $B_{n}^\dag$ or $A_n^\dag$ that create an extra node. Thus, the overall number of the nodes does not change. In addition, the normalization does not require positive semi-definiteness of the energy eigenvalues, as in the standard case. This is good because negative energy states appear when $n>0$. 

For any $n$ there is always one missing state $k=n$ which can be obtained by solving the first order differential equation $B_{n}\tilde\psi_n^{(n)}=0$ (by construction the state $\tilde\psi_n^{(n)}$ has to be annihilated by the operator $B_{n}$):
\begin{eqnarray*}{d\tilde\psi_n^{(n)}(x)\over dx}&=&-\left(W_n(x)+\frac{\psi_n^2(x)}{C +\int_{x_0}^x{\psi_n^2(s)ds}}\right) \tilde\psi_n^{(n)}(x)\\&=&{d\over dx}\left(\ln\frac{\psi_n}{C +\int_{x_0}^x{\psi_n^2(s)ds}}\right)\tilde\psi_n^{(n)}(x)\end{eqnarray*}
Therefore, \begin{equation}\tilde \psi_n^{(n)}(x)=N(C)\times\frac{\psi_n}{C +\int_{x_0}^x{\psi_n^2(s)ds}}\end{equation}
with the corresponding energy $\tilde E_n^{(n)}=0$.  All other energy eigenvalues satisfy $\tilde E_k^{(n)}=E_k-E_n$. The normalization constant $N(C)$ depends on the parameter $C$ and other parameters of the potential such as width, depth etc. It is a constraint that allows one to determine the values of $C$ for which the potentials $\tilde V_-^{(n)}(x)$  are nonsingular and eigenfunctions $\tilde\psi_k^{(n)}(x)$ are well-defined. 

One observes that the intertwining relationship between the Hamiltonians $H_-^{(n)}$ and $\tilde H_-^{(n)}$ is of the second order: \[\tilde H_-^{(n)}B_n^\dag A_n=B_n^\dag A_n H_-^{(n)} \mbox{ and } H_-^{(n)}A_n^\dag B_n=A_n^\dag B_n\tilde H_-^{(n)}.\] In the second-order SUSY QM \cite{Hernandez} two different Hamiltonians are intertwined by an operator of the second-order in derivatives, say, $A=\partial_x^2+\eta(x)\partial_x+\gamma(x)$. If $A$ can be written as a product of two first-order differential operators with real superpotentials, then we call it reducible (otherwise one refers to it as irreducible). Thus, our construction is equivalent to the second-order SUSY QM with the reducible operator $A=-B_n^\dag A_n$. Performing an explicit factorization one finds that $-\eta(x)=f_n(x)$ and $-\gamma(x)=V_-^{(n)}(x)+f_n(x)W_n(x)$. Pros and cons of these related approaches are discussed in detail in the concluding section.   

From now on we will discuss the degeneracy of energy levels of the Hamiltonians $H_-^{(n)}$ and $\tilde H_-^{(n)}$ only, leaving aside the Hamiltonian $H_+^{(n)}$ which plays an intermediate role in this construction.   

\section{Examples}
 Here we illustrate the results developed in the preceding section by providing examples that arise from well-known potentials and obtain some previously unreported potentials which might be of interest in various fields of physics and chemistry. One can also consult the ref.~\cite{Berger} where factorizations of the harmonic oscillator potential were performed.

\subsection{Morse potential}
Let us first consider the Morse potential \begin{equation}V_-^{(0)}(x)=A^2-B(2A+\alpha)e^{-\alpha x}+B^2e^{-2\alpha x}\end{equation} where the constants $A, B$ and $\alpha$ are nonnegative. There is a finite number of energy levels $E_k=k\alpha(2A-k\alpha)$ where $k$ takes integer values from zero to the greatest value for which $k\alpha< A$. For concreteness let us take $A=2$ and $\alpha=B=1$. The partner potential $\tilde V_-^{(0)}(x)$ is obtained from the ground state wave function $\psi_0(x)=e^{-2x -e^{-x}}$ of the potential $V_-^{(0)}(x)=4-5e^{-x}+e^{-2 x}$:
\begin{eqnarray*}&&\tilde V_-^{(0)}(x)=4-5e^{-x}+e^{-2 x}\\&-&16{d\over dx}\left(\frac{e^{-4x-2e^{-x}}}{C+e^{-2e^{-x}}(3+6e^{-x}+6e^{-2x}+4e^{-3x})}\right).\end{eqnarray*}

As the potential $V_-^{(0)}(x)$ it has only two bound states with eigenvalues $\tilde E_0^{(0)}=0$ and $\tilde E_1^{(0)}=3$. The normalized ground state wave function is \[\tilde\psi_0^{(0)}(x)=\frac{\sqrt{8C(C+3)\over3} \,e^{-2x -e^{-x}}}{C+e^{-2e^{-x}}(3+6e^{-x}+6e^{-2x}+4e^{-3x})}.\] Hence, the potential $\tilde V_-^{(0)}(x)$ is nonsingular as long as $C\not\in[-3,0]$ (see Fig. 1). 
\begin{figure}[h] 
  \centering
  \includegraphics[bb=0 0 240 235,width=3.2in,height=3.2in,keepaspectratio]{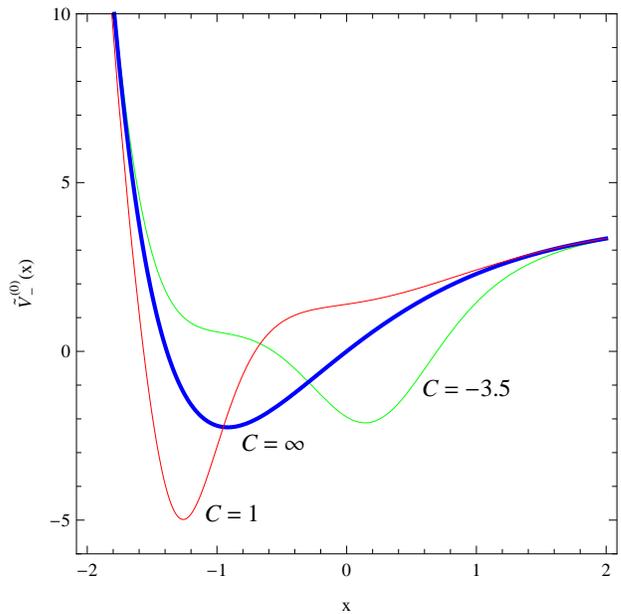}
  \caption{A few members of the one-parameter family of potentials $\tilde V_-^{(0)}(x)$ isospectral to the Morse potential $V_-^{(0)}(x)$ with $A=2$ and $\alpha=B=1$ (thick blue line). }
  \label{fig:fig.1}
\end{figure}

The normalized wave function $\tilde\psi_1^{(0)}(x)$ is determined by applying the operator $B_0^\dag A_0$ to the first (and only) normalized excited state $\psi_1(x)=2/\sqrt3e^{-x-e^{-x}}(3-2e^{-x})$ of the potential $V_-^{(0)}(x)$:\[\tilde\psi_1^{(0)}(x)=\frac{2e^{-e^{-x}} (6 + 12 e^x + 9 e^{2 x}) +Ce^{e^{-x}}( 3 e^{2x} - 2 e^{x})}{\sqrt3(4 + 6 e^x + 6 e^{2 x} + 3 e^{3 x} +  C e^{2 e^{-x} + 3 x})}.\]

We would like to remind the ladder operators for the wave functions of the Morse potential given in (5) and explicitly derive them for the wave functions of the isospectral partner potential. Let's denote $s=A/\alpha$ and $y=2B/\alpha e^{-\alpha x}$ which is the common choice in the SUSY QM literature. Then for the creation $K_+$ and annihilation $K_-$ operators we have \cite{Dong}:     
\[K_+=\left[\partial_y+{{s-n}\over y}-{s+1/2\over {2s-2n-1}}\right]\] and  \[K_-=-\left[\partial_y-{{s-n}\over y}+{s+1/2\over {2s-2n+1}}\right]\] (we note that $K_-\ne K_+^\dag$) with the following effect $K_+\psi_k(y)\propto\psi_{k+1}(y)$ and $K_-\psi_{k+1}(y)\propto\psi_{k}(y)$. The proportionality factors can be calculated after normalizing the eigenfunctions $\psi_k(y)=y^{s-k}e^{-y/2}L_n^{2s-2k}(y)$ where $L_k^{2s-2k}(y)$ are associated Laguerre polynomials. 

The equation (3) enables us to deduce the ladder operators for the eigenvectors $\tilde \psi_k^{(n)}(y)$ of the potential $\tilde V_-^{(n)}(x)$ whose energy spectrum is identical to that of the Morse potential $V_-^{(n)}(x)$. The corresponding raising and lowering operators for $\tilde \psi_k^{(n)}(y)$ with $k\ne n$ are $(B_n^\dag A_n)K_+(A_n^\dag B_n)$ and $(B_n^\dag A_n)K_-(A_n^\dag B_n)$. Exploration of the higher-order ladder operators is the direct consequence of extending the first-order SUSY QM.

\subsection{CPRS potential}
In ref. \cite{Carinena} Cari\~nena, Perelomov, Ra\~nada and Santander (CPRS) have studied the following one-dimensional non-polynomial exactly solvable potential (we define our Hamiltonian to be $H_-^{(0)}=2H_{\text{CRPS}}+3$): \[V_-^{(0)}(x)=x^2+3+8\frac{2x^2-1}{(2x^2+1)^2}.\]
This potential asymptotically behaves like a simple harmonic oscillator but its minimum at the origin is much deeper than in case of the harmonic oscillator. Using SUSY QM techniques it was shown by Fellows and Smith \cite{Fellows} that $V_-^{(0)}(x)$ is a partner potential of the harmonic oscillator $x^2+5$ and, therefore, their energy levels are the same. Here we further analyze the CPRS potential and find new potentials with the oscillator spectrum (see also ref. \cite{Berger}).

The ground state energy $E_0=0$ and wave function \[\psi_0(x)=\frac{e^{-x^2/2}}{2x^2+1}\]of the potential $V_-^{(0)}(x)$
allows one to find  its isospectral partner \begin{eqnarray*}&&\tilde V_-^{(0)}(x)=x^2+3+8\frac{2x^2-1}{(2x^2+1)^2}\\&-8&{d\over dx}\left(\frac{e^{-x^2}}{2x(2x^2+1)e^{-x^2}+(2x^2+1)^2(C+\sqrt\pi\operatorname{erf} x)}\right)\end{eqnarray*} which has no singularities when $|C|>\sqrt\pi$ (see Fig. 2) as follows from normalizing the ground state wave function $\tilde\psi_0^{(0)}(x)$ (see below). 
\begin{figure}[h] 
  \centering
  \includegraphics[bb=0 0 240 235,width=3.2in,height=3.2in,keepaspectratio]{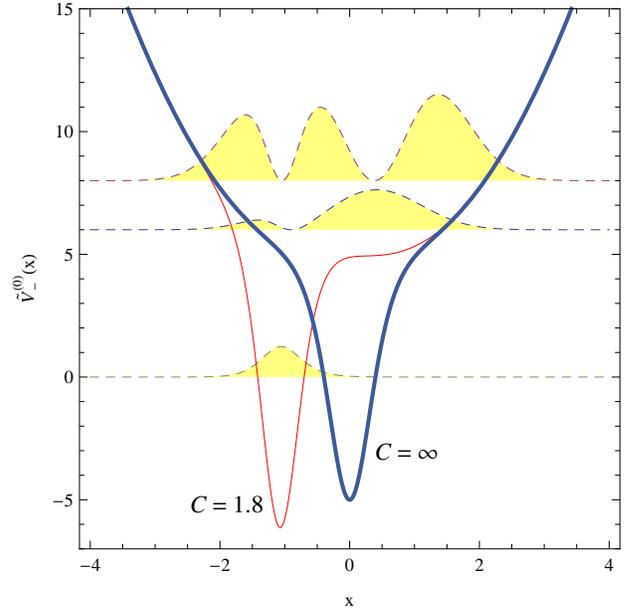}
  \caption{Plot of the potential $\tilde V_-^{(0)}(x)$ with $C=1.8$ (close to $\sqrt\pi$) and the unnormalized probability densities (dashed line at the corresponding level position) for its three lowest energy levels. The limit $C\to\infty$ corresponds to the CPRS potential (thick blue line).}
  \label{fig:fig.2}
\end{figure}

Its eigenvalues are the same as that of the potential $V_-^{(0)}(x)$ and given by $\tilde E_k^{(0)}=2k+4$ for $k=1,2,\hdots$. The normalized ground state wave function \[\tilde\psi_0^{(0)}(x)=\frac{\sqrt{2(C^2-\pi)/\sqrt\pi}\,e^{-x^2/2}}{2xe^{x^2}+(2x^2+1)(C+\sqrt\pi\operatorname{erf}(x))}\] corresponds to the energy eigenvalue $\tilde E_0^{(0)}=0$. The rest of the eigenfunctions can be derived using equation (3).

Neither Cari\~nena et al., nor Fellows and Smith provided the raising and lowering operators for the wave functions $\psi_k(x)$ of the CPRS potential. Here we address the question of finding ladder operators for the CPRS potential and its isospectral partner. Taking into account that the CPRS potential itself is a partner of the harmonic oscillator, we obtain its raising $A^\dag a^\dag A$ and lowering $A^\dag a A$ operators where \[A=\partial_x+x+\frac{4x}{2x^2+1}\] is needed to move between the CPRS potential and harmonic oscillator whose creation and annihilation operators are $a^\dag$ and $a$ respectively. Thus, the ladder operators for the wave functions $\tilde\psi_k^{(n)}(x)$ of the potential $\tilde V_-^{(n)}$ become $(B_n^\dag A_n)A^\dag a^\dag A(A_n^\dag B_n)$ and $(B_n^\dag A_n)A^\dag a A(A_n^\dag B_n)$ for $k\ne n$.

\subsection{Infinite square well potential}
Despite its simplicity, the one-dimensional infinite square well potential with a deformed bottom requires some new techniques for obtaining solutions of the corresponding Schr\"{o}dinger equation and usually one is unable to solve it exactly. In a recent paper \cite{Alhaidari}, exact solution for the problem with sinusoidal bottom has been deduced. In this subsection we explicitly find potentials with undulating bottom and energy spectrum coinciding with that of the infinite square well.
 
The wave functions and energy eigenfunctions of the infinite square well potential $V_-^{(0)}(x)=-\pi^2/L^2$ of width $L$ are given by $\psi_k(x)=\sin({(k+1)\pi x/ L})$ with $0\le x\le L$ and $E_k=k(k+2)\pi^2/L^2$. Using this time for diverseness the first excited state wave function $\psi_1(x)$ we find a pair of partner potentials, namely, the infinite square well potential with flat bottom \[V_-^{(1)}(x)=-4\pi^2/L^2\] and the infinite square well potential with non-flat bottom also defined in the region $0\le x\le L$ (see Fig. 3):
\begin{figure}[h] 
  \centering
  \includegraphics[bb=0 0 240 235,width=3.2in,height=3.2in,keepaspectratio]{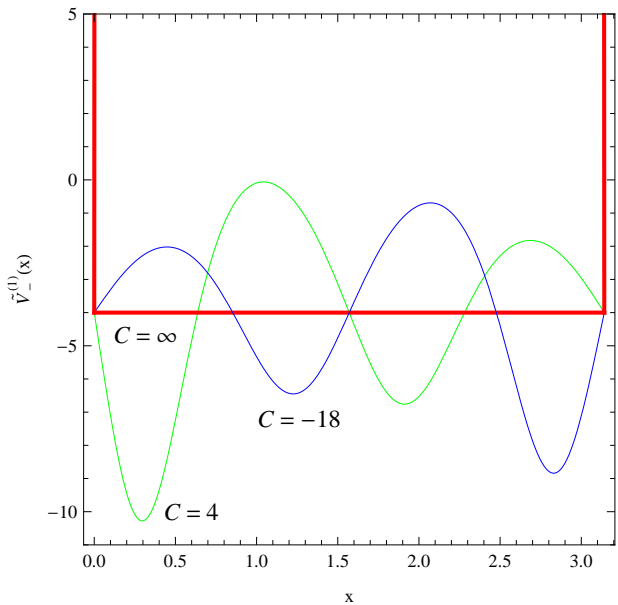}
  \caption{Selected members of the family of one-parameter potentials $\tilde V_-^{(1)}(x)$. The limit $C\to\infty$ corresponds to the infinite square well $V_-^{(0)}(x)=-4$ of width $L=\pi$ (thick red line).}
  \label{fig:fig.3}
\end{figure}

\[\tilde V_-^{(1)}(x)=-{4\pi^2\over L^2}-16{d\over dx}\left(\frac{\sin^2{(2\pi x/ L)}}{C+4 x-L/\pi\sin{(4\pi x/ L)}}\right).\] Both of the potentials have identical energy spectra $\tilde E_k^{(1)}=(k-1)(k+3)\pi^2/L^2$. The normalized first excited state of the potential $\tilde V_-^{(1)}(x)$ is calculated from (4) and reads \[\tilde\psi_1^{(1)}(x)=\sqrt{2C(C+4L)\over L}\frac{\sin{(2\pi x/ L)}}{C+4 x-L/\pi\sin{(4\pi x/ L)}}\] provided that $C\not\in[-4L,0]$. The wave functions $\tilde\psi_0^{(1)}(x)$, $\tilde\psi_2^{(1)}(x),\hdots$ can be found from (3). We only calculate the normalized lowest state eigenfunction: \[\tilde\psi_0^{(1)}(x)=\frac{\sin{\pi x\over L}\left(3\pi(C+4x)-8L\sin{2\pi x\over L}+L\sin{4\pi x\over L}\right)}{3\sqrt{L\over 2}\left(L\sin{4\pi x\over L}-\pi(C+4\pi)\right)}.\] It corresponds to the negative energy $\tilde E_0^{(1)}=-3\pi^2/L^2$ as expected since the potential $\tilde V_-^{(1)}(x)$ is generated by the first excited state of the original potential. 
Note that the potential $\tilde V_-^{(1)}(x)$ satisfies \[\tilde V_-^{(1)}(C, x)=\tilde V_-^{(1)}(C+2L, x+L/2).\] 
 
It is known \cite{Dong} that the eigenvectors $\psi_k(x)$ of the Hamiltonian $H_-^{(n)}$ admit the following creation and annihilation operators: \[M_+=\cos{\left(\pi x\over L\right)}\hat k+ {L\over\pi}\sin{\left(\pi x\over L\right)}\partial_x\] and \[M_-=\left[\cos{\left(\pi x\over L\right)}\hat k- {L\over\pi}\sin{\left(\pi x\over L\right)}\partial_x\right]\hat k^{-1}(\hat k-1)\] where one defines the "number" operator $\hat k$ and its inverse $\hat k^{-1}$ such that  $\hat k \psi_k(x)=(k+1)\psi_k(x)$ and $\hat k^{-1}\psi_k(x)=(k+1)^{-1}\psi_k(x)$. The ladder operators $M_\pm$ obey \[M_-\psi_k(x)=k\psi_{k-1}(x) \mbox{ and } M_+\psi_k(x)=(k+1)\psi_{k+1}(x).\] 
 
It is not hard to convince yourself that the raising and lowering operators for the wave functions $\tilde \psi_k^{(n)}$ of the partner isospectral Hamiltonian $\tilde H_-^{(n)}$ are given by $(B_n^\dag A_n) M_+(A_n^\dag B_n)$ and $(B_n^\dag A_n) M_-(A_n^\dag B_n)$ respectively for $k\ne n$ (when $k=n$ use equation (4)). 

\subsection{Two-parameter set of potentials isospectral to the harmonic oscillator}
Given an eigenfunction $\psi_n(x)$ of the potential $V_-^{(0)}(x)$ one can find the wave function $\tilde\psi_k^{(n)}(x)$ of the one-parameter potential $\tilde V_-^{(n)}$ using the equation (3). Now one can repeat this procedure and instead of the eigenfunction $\psi_n(x)$ in (2) and (3) use $\tilde\psi_k^{(n)}(x)$ to obtain a two-parameter potential $\tilde V_-^{(n,k)}(x)$ and its eigenfunctions. One can go on with this construction and obtain well defined multi-parameter potentials strictly isospectral to the potential $V_-^{(k)}(x)$. 

Let us focus on the harmonic oscillator $V_-^{(0)}(x)=x^2-1$ (with $\omega=2$)  whose ground state wave functions is $\psi_0(x)=e^{-x^2/2}$. The potential $\tilde V_-^{(0)}(x)$ is carefully discussed in refs. \cite{Mielnik, Berger, Abraham} each using different approaches, so in the following we omit unnecessary calculations and only state its normalized first excited state wave function: \[\tilde\psi_1^{(0)}(x)=\sqrt{2\over\sqrt\pi}\frac{e^{-3x^2/2}(1+2C x e^{x^2}+\sqrt\pi xe^{x^2}\operatorname{erf}(x))}{2C+\sqrt\pi\operatorname{erf}(x)}\] where $|C|>\sqrt\pi/2$ to guarantee non-singularity of the potential $\tilde V_-^{(0)}(x)$. Applying (2) to the wave function $\tilde\psi_1^{(0)}(x)$ we get the two-parameter potential (see Fig. 4): \begin{eqnarray*}&&\tilde V_-^{(0,1)}(x)=x^2-3\\&-&2{d\over dx}\left(\frac{e^{-x^2}}{C+\sqrt\pi/2\operatorname{erf}(x)}+\frac{(\tilde\psi_1^{(0)}(x))^2}{\tilde C+\int_{x_0}^x{(\tilde\psi_1^{(0)}(s))^2}ds}\right)\end{eqnarray*} which is isospectral to the potential $\tilde V_-^{(0)}(x)-2$, which is in turn isospectral to the harmonic oscillator $V_-^{(1)}(x)=x^2-3$, i.e. its energy levels are $\tilde E_k^{(0,1)}=2(k-1)$. 

The potential $\tilde V_-^{(0,1)}(x)$ is non-singular for any $C\ne0$ and $|\tilde C +1/(4C)|>\sqrt\pi/4$ as follows from normalizing its ground state wave function. This family includes the oscillator potential $x^2-3$ in the limit $C, \tilde C \to\infty $; the potential $\tilde V_-^{(0)}(x)$ arises when $\tilde C\to\infty$; and finally $\tilde V_-^{(0,1)}(x)$ reduces to the potential $\tilde V_-^{(1)}(x)$ \cite{Berger} in the limit $C\to\infty$.  
 
\begin{figure}[h] 
  \centering
  \includegraphics[bb=0 0 240 235,width=3.2in,height=3.2in,keepaspectratio]{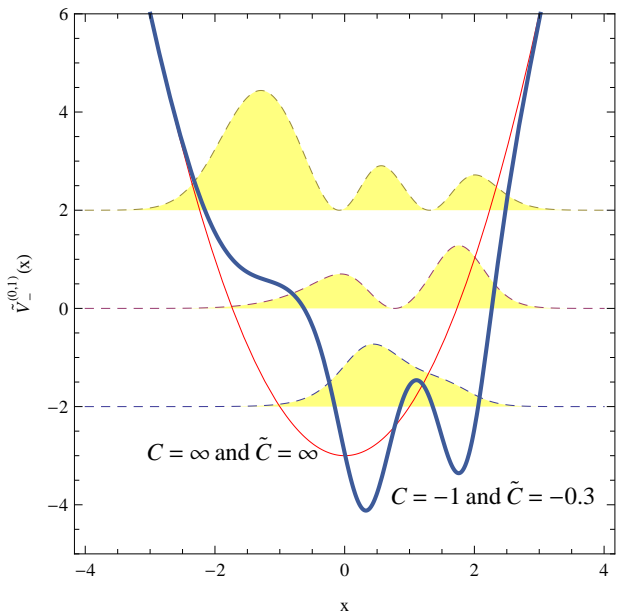}
  \caption{Plot of the potentials $\tilde V_-^{(0,1)}(x)$, $V_-^{(1)}(x)=x^2-3$ and the non-normalized probability densities (dashed line at the corresponding level position) for the three lowest energy levels of $\tilde V_-^{(0,1)}(x)$.}
  \label{fig:fig.4}
\end{figure}
Let's briefly mention how to obtain its eigenfunctions. There is an expression similar to (3) for $k=2,3,\dots$: \[\tilde\psi_k^{(0,1)}(x)\propto\tilde B_1^\dag \tilde A_1\tilde \psi_k^{(0)}(x)\propto\tilde B_1^\dag \tilde A_1B_0^\dag A_0\psi_k(x)\] where $\tilde \psi_k^{(0)}(x)$ and $\psi_k(x)$ are the eigenfunctions of the potential $\tilde V_-^{(0)}(x)$ and the harmonic oscillator accordingly. The operators $\tilde B_1^\dag$, $\tilde A_1$ are defined by \[\tilde A_1=\partial_x-\partial_x\ln\tilde \psi_1^{(0)}(x)\] and \[\tilde B_1^\dag=-\partial_x+\partial_x\ln\frac{\left(\tilde C+\int_{x_0}^x{(\tilde\psi_1^{(0)}(s))^2}ds\right)}{\tilde \psi_1^{(0)}(x)}.\] Lastly, the raising and lowering operators for the eigenvectors $\tilde\psi_k^{(0,1)}(x)$ are given by $\tilde B_1^\dag\tilde A_1B_0^\dag A_0A_0^\dag A_0^\dag B_0\tilde A_1^\dag \tilde B_1$ and $\tilde B_1^\dag\tilde A_1B_0^\dag A_0A_0A_0^\dag B_0\tilde A_1^\dag \tilde B_1$ with $A_0^\dag$, $A_0$  being the creation and annihilation operators of the  harmonic oscillator.
 
The two-parameter family of potentials with oscillator spectrum was also derived by the so-called second order intertwining technique in \cite{Fernandezz}. The advantages of the presented technique of getting multi-parameter sets of isospectral potentials are apparent.  
\section{Conclusion}
After the discovery of supersymmetry in string theory and then field theory, factorization was recognized as the application of supersymmetry to quantum mechanics. The non-uniqueness of factorization serves as an avenue for the construction of many isospectral potentials. In this paper, we have explored the generality of this method by extending it to the excited states of a potential. Some nonsingular isospectral potentials that arise from the technique have been presented in this paper.
These include one-parameter extensions of the well-known infinite square well and Morse potentials as well as not so familiar CPRS potential and two-parameter extension of the harmonic oscillator.  For some potentials the associated wave functions and probability densities have been derived and plotted. The ladder operators were determined explicitly. The application of this technique may be of significant interest because it can be applied to any one-dimensional quantum mechanical potential. 

The most general approach in the second-order SUSY QM is based on an arbitrary solution of the Schr\"odinger equation for the initial potential, rather than on its ground or excited state wave functions as discussed in the present article. However, there are certain advantages in such a presentation. For example, one can explicitly construct the ladder operators for both isospectral Hamiltonians. It is also possible to avoid some technical complexities of the most general approach by mimicking the traditional first-order SUSY QM. For instance, in the second-order SUSY QM none of the expressions $AA^\dag$ or $A^\dag A$ coincide with any of the isospectral partner Hamiltonians, but are quadratic forms in them. For comparison in our construction, which is based on the non-uniqueness of factorization, the appearance of the atypical Hamiltonian $H_+^{(n)}$ at the intermediate stage does not affect the isospectral partner Hamiltonians $\tilde H_-^{(n)}$ and $H_-^{(n)}$.

\section*{Acknowledgments}
N.U. was assisted by the Hutton Honors College Research Grant. M.B was supported in part by the U.S.
Department of Energy under Grant No.~DE-FG02-91ER40661. 

\end{document}